\documentclass[12pt]{article}
\usepackage{epsfig}
\usepackage{axodraw}

\newcommand{\xF}{x_{\mathrm{F}}}
\newcommand{\pt}{p_{\perp}}
\newcommand{\kt}{k_{\perp}}
\renewcommand{\c}{\mathrm{c}}
\renewcommand{\d}{\mathrm{d}}

\newcommand{\p}{\mathrm{p}}
\newcommand{\q}{\mathrm{q}}
\renewcommand{\u}{\mathrm{u}}
\renewcommand{\d}{\mathrm{d}}

\newcommand{\D}{\mathrm{D}}

\renewcommand{\B}{\mathrm{B}}

\newcommand{\cbar}{\overline{\mathrm{c}}}

\newcommand{\qbar}{\overline{\mathrm{q}}}
\newcommand{\ubar}{\overline{\mathrm{u}}}
\newcommand{\pbar}{\overline{\mathrm{p}}}
\newcommand{\Bbar}{\overline{\mathrm{B}}}

\newcommand{\Py}{{\sc{Pythia}}}

{\end{list}}
\newcounter{enumct}

\newlength{\abstwidth}
\begin{document}

\sloppy

\pagestyle{empty}

\begin{flushright}
LU TP 99--28\\
September 1999
\end{flushright}
 
\vspace{\fill}

\begin{center}
{\LARGE\bf Heavy Quark Production Asymmetries}\\[10mm]
{\Large E. Norrbin\footnote{emanuel@thep.lu.se}} \\[3mm]
{\it Department of Theoretical Physics,}\\[1mm]
{\it Lund University, Lund, Sweden}
\end{center}
 
\vspace{\fill}
 
\begin{center}
{\bf Abstract}\\[2ex]
\begin{minipage}{\abstwidth}
In the hadroproduction of charm (or heavy flavours in general) in the context of string
fragmentation, the pull of a beam remnant at the other end of a string
may give a charm hadron more energy than the perturbatively produced one. The collapse
of a low-mass string to a single hadron is the extreme case in this direction, and
gives rise to asymmetries between leading and non-leading charm hadrons. We study these
phenomena within the Lund string fragmentation model and improve
the modelling in part by a consideration of hadroproduction data. Applications include
heavy quark production in any collision between hadron-like particles such as
$\gamma\mathrm{p}$ at HERA and pp at HERA-B or the LHC.

\end{minipage}
\end{center}
 
\vspace{\fill}

\clearpage
\pagestyle{plain}
\setcounter{page}{1}

\section{Introduction}
Several experiments at fixed-target energies have observed asymmetries between leading
and non-leading charmed hadrons in $\pi^-\p$ collisions  \cite{obs}. In \cite{ENTS}
we study this effect in the context of the Lund string fragmentation model \cite{AGIS} and find
that it is possible to improve the model to obtain a good description of the asymmetry data.
In this model, collapses of light colour singlets mainly account for the asymmetry.

We also study single charm spectra and particle/anti-particle
correlations and find good agreement between the model and single charm spectra, but
fail to describe the correlation data from E791 \cite{hq98proc}. This is in
contrast to NLO calculations \cite{NLO}, which adequately describe single charm
spectra for non-leading particles and some correlations, but fail to describe the
asymmetry and leading-particle single charm spectra.

A related effect included in the string model is the non-perturbative 'beam drag'.
In this model the produced heavy quarks are connected to the beam remnants
by colour singlet strings. Through soft interactions with the beam remnant the charm quark
can gain energy and momentum to produce a heavy hadron at a larger rapidity than
that of the heavy quark.

\section{Model aspects}

\begin{figure}
\begin{center}
\mbox{\epsfig{file=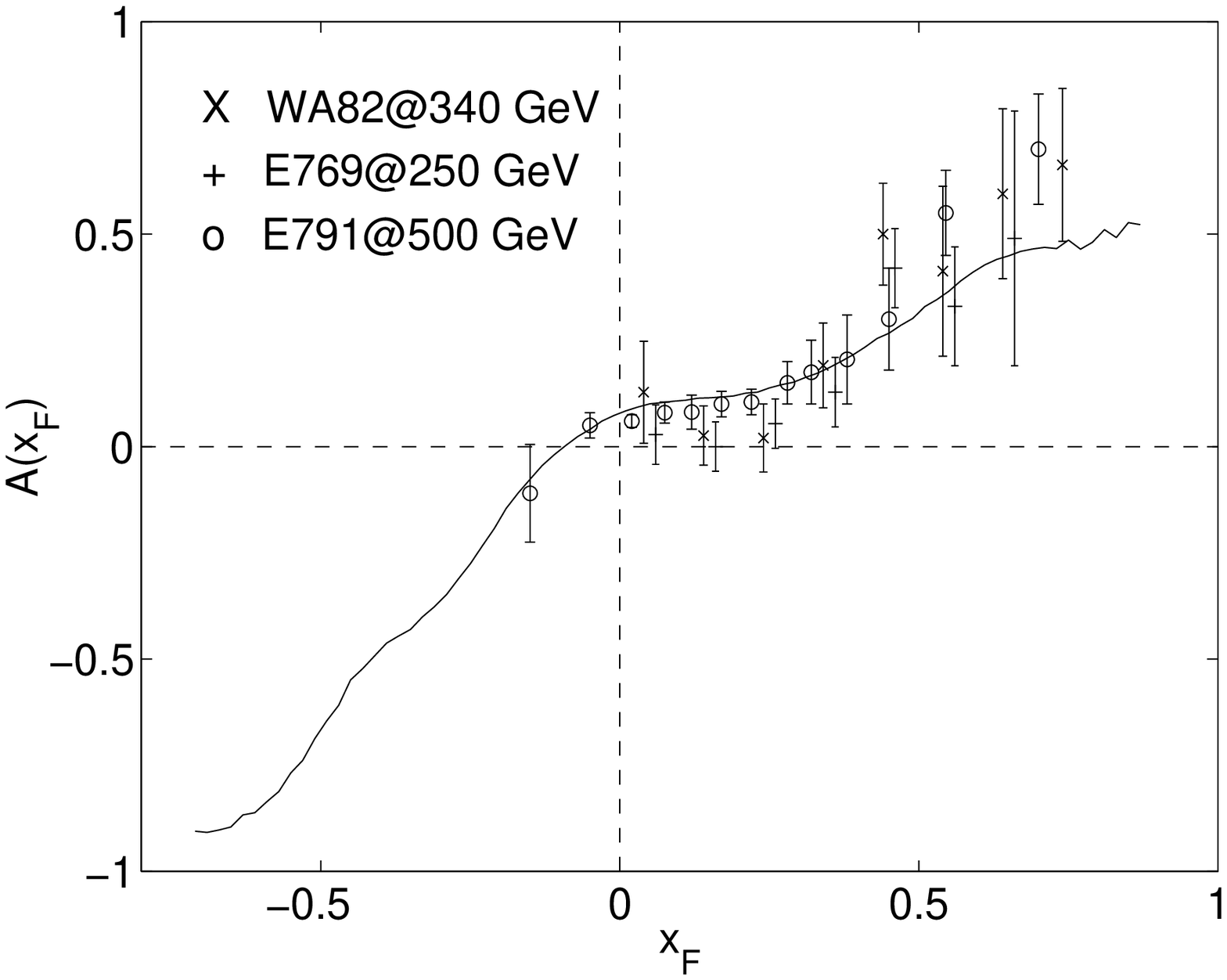, width=79mm}}
\mbox{\epsfig{file=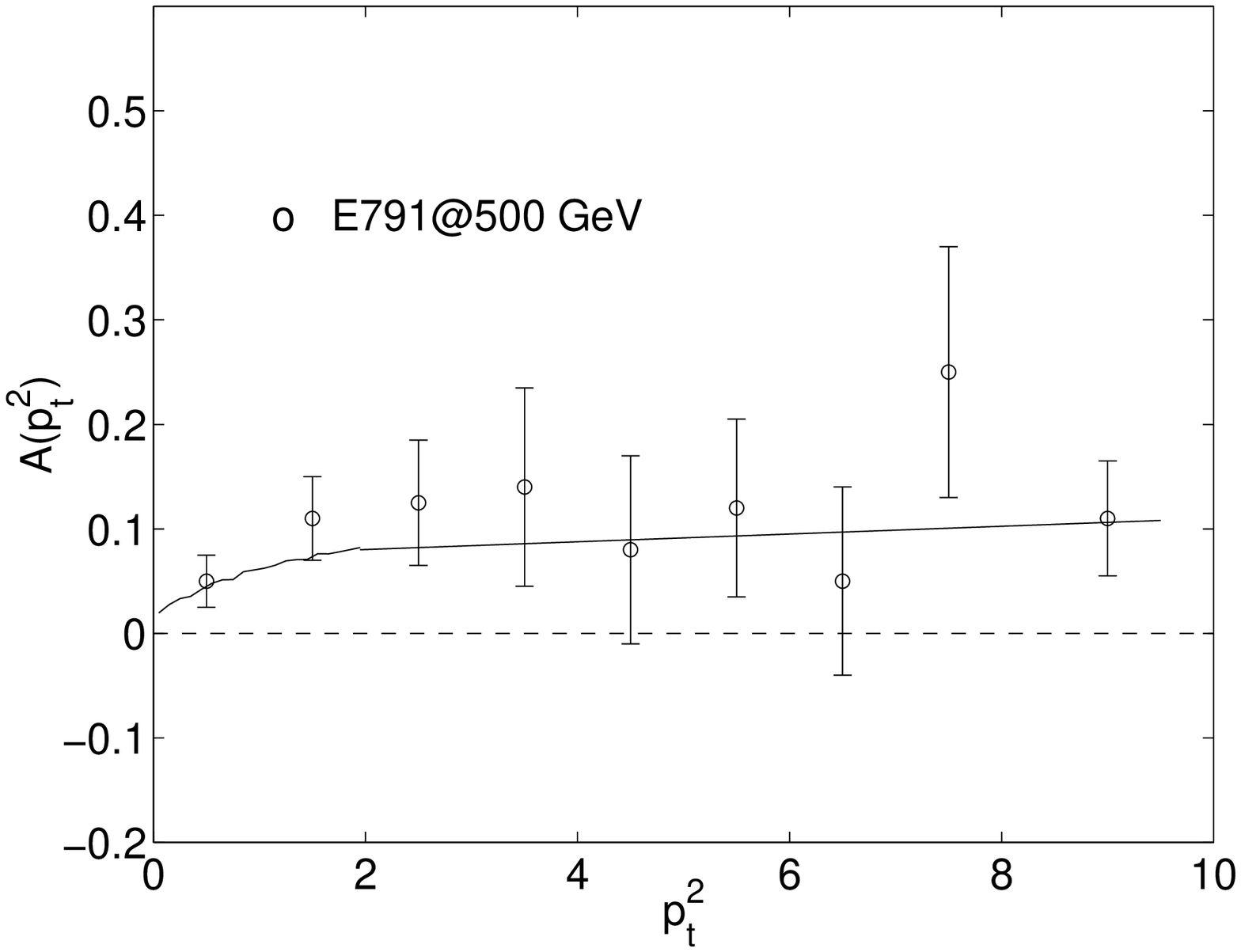, width=79mm}}
\end{center}
\caption[]{Experimental results on asymmetries compared to the modified version of
\Py~6.1 described in this note. Data is from $\pi^-\p$ fixed-target experiments,
where $\xF > 0$ is the pion fragmentation region.}
\label{fig:exp}
\end{figure}

\begin{figure}
\begin{center}
\begin{picture}(135,75)(0,0)
\SetOffset(0,-15)

\small
\Text(12,30)[r]{$\p^+$}
\Text(12,70)[r]{$\pi^-$}
\Line(15,30)(25,30)
\Line(15,70)(25,70)
\GOval(30,30)(15,5)(0){0.5}
\GOval(30,70)(15,5)(0){0.5}

\Text(51,35)[b]{$\u$}
\Text(51,67)[t]{$\ubar$}
\Line(35,35)(60,50)
\Line(35,65)(60,50)
\Gluon(60,50)(90,50){5}{4}
\Line(90,50)(115,35)
\Line(90,50)(115,65)
\Text(120,33)[l]{$\c$}
\Text(119,68)[l]{$\cbar$}

\Line(35,24)(115,24)
\Line(35,79)(115,79)
\Text(119,25)[l]{$\u\d$}
\Text(119,80)[l]{$\d$}

\put(127,60){\oval(4,10)[r]}
\put(131,14){\oval(4,10)[r]}

\end{picture}

\end{center}
\caption{String configuration in a $\pi^-\p$ collision
with quark fusion charm production.}
\label{fig:strings}
\end{figure}
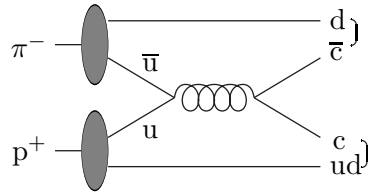

\begin{figure}
\begin{center}
\mbox{\epsfig{file=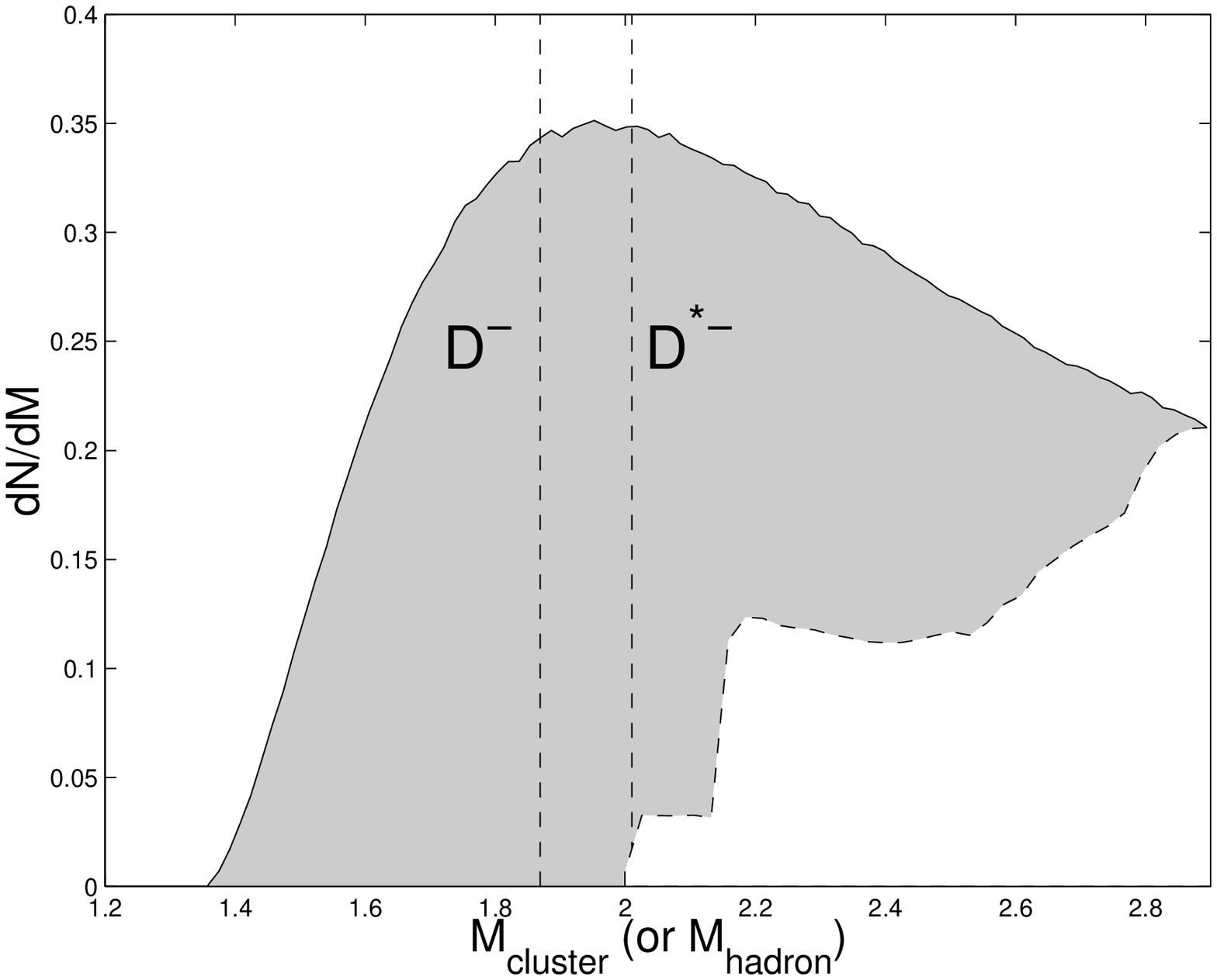, width=79mm}}
\mbox{\epsfig{file=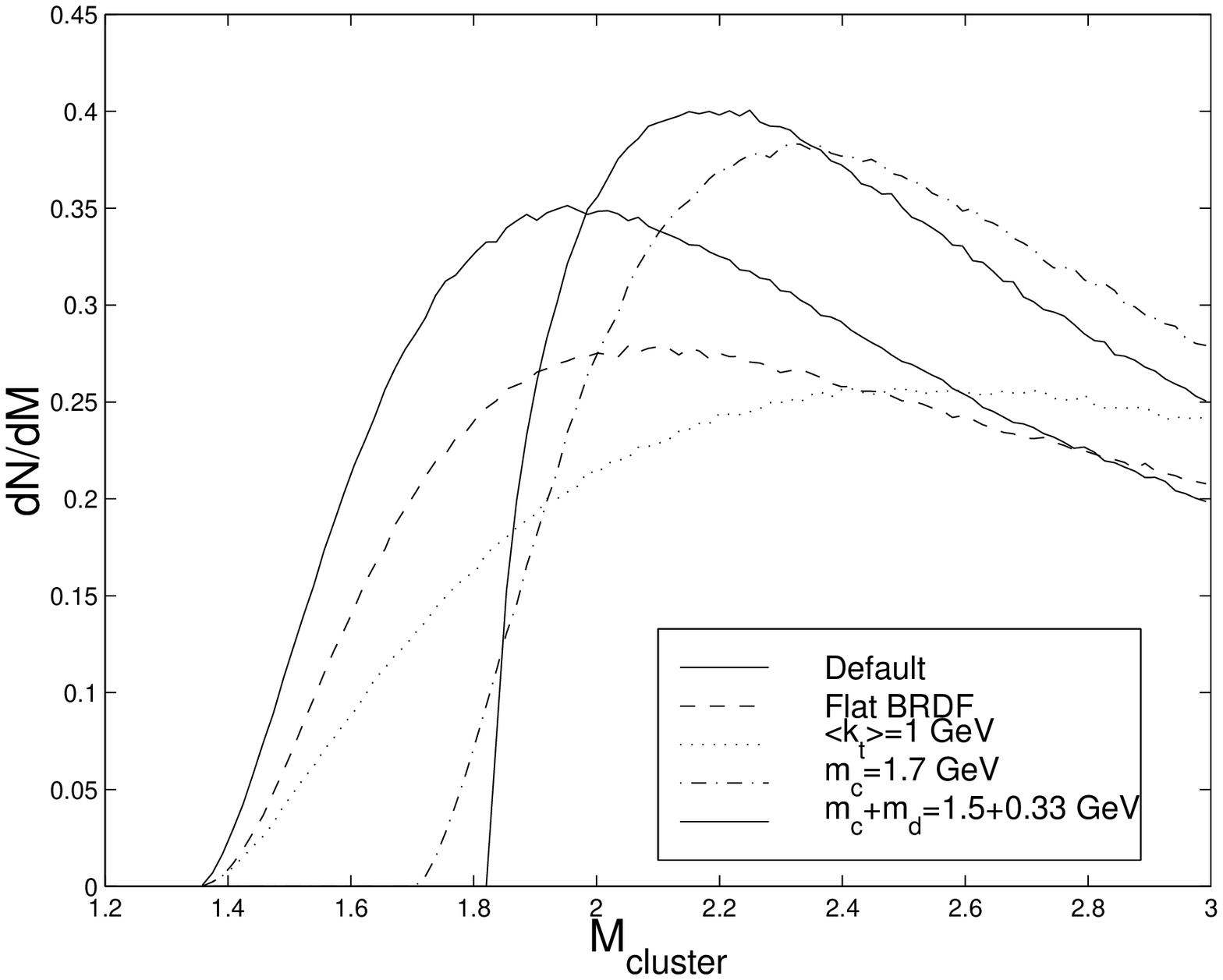, width=79mm}}
\end{center}
\caption{Distribution of $\cbar$-d colour singlet masses in a $\pi^-\p$ collision using different
parameters of the model. The dashed curve in the first figure is the distribution of
invariant mass of the hadronic system produced from the colour singlet string or cluster.
Below the two-particle threshold only the one-particle states $\D^-$ and $\D^{*-}$ are available.}
\label{fig:massdist}
\end{figure}

The asymmetry between $\D^+$ and $\D^-$ is described by the asymmetry parameter
\begin{equation}
A = A(\xF,\pt)= \frac{\sigma(\D^-)-\sigma(\D^+)}{\sigma(\D^-)+\sigma(\D^+)} ~,
\end{equation}
which is observed to be an increasing function of $\xF$ in $\pi^-\p$ collisions.
In fig.~\ref{fig:exp} experimental results are compared to a modified version
of the string model as implemented in the \Py~6.1 \cite{Pythia} event generator.
In the Lund string fragmentation model the produced charm quarks are connected to
the beam remnants by colour singlet strings as e.g. in fig.~\ref{fig:strings}.
The colour of the incoming anti-up quark is inherited by the outgoing anti-charm quark,
which forms a colour singlet together with the pion beam remnant
down quark. The distribution of $\cbar$-$\d$ string masses near threshold is shown in
fig.~\ref{fig:massdist}, both using the default parameters and with some modifications.

Depending on the mass of a string, different hadronization mechanisms are used:
\begin{description}
\item[Large mass]
($M_{\mathrm{string}} > m_{\mathrm{c}} + m_{\mathrm{q}} + 1~\mathrm{GeV}$)
Ordinary string fragmentation with a continuum of phase-space states.
\item[Low mass]
The colour-singlet system forms a cluster that undergoes two-body decay into
two hadrons.
\item[Very low mass]
If the invariant mass of a cluster is so small that two-body decay is not possible,
the cluster is forced to collapse into a single hadron with compatible flavour
contents. This gives rise to an asymmetry favouring leading particles.
\end{description}

The qualitative nature of the asymmetry is thus described by the string model.
However, the quantitative behaviour depends on the parameters of the model.
The model aspects considered are described in
more detail in \cite{ENTS}, so here we just give a brief summary:

\begin{itemize}
\item
The masses used for the  light and heavy quarks regulate the fraction of collapses. More
collapses gives rise to a larger asymmetry. We use $m_{\mathrm{c}} + m_{\mathrm{q}} = 1.5~\mathrm{GeV} + 0.33~\mathrm{GeV}$, i.e. constituent masses and the charm mass that
gives the best fit to the total charm cross section.
\item
Width of the primordial $\kt$ distribution. A larger intrinsic $\kt$ of the partons
entering the hard interaction gives rise to an asymmetry also at larger $\pt$ which
is in agreement with data. We use a Gaussian with $\langle \kt \rangle = 1~\mathrm{GeV}$.
\item
Beam remnant distribution functions. The energy-momentum of the
beam remnant has to be split between the constituent quarks. This aspect is varied by using
different beam remnant distribution functions parameterizing the fraction of
energy-momentum taken by the constituents. We use an intermediate scenario.
\item
Threshold behaviour between collapse and decay of a cluster. The transition between
cluster collapse and cluster decay as the cluster mass is increased can be slow or
fast. We use an intermediate scenario.
\item
Energy conservation in collapse. When a string with a very low mass collapses into
a hadron, both energy and momentum cannot be conserved simultaneously. We consider
different ways to shuffle energy between strings and partons in an event, but find
that observable quantities are not sensitive to this aspect.
\end{itemize}

\section{Beam drag}

\begin{figure}
\begin{center}
\begin{picture}(170,50)(0,0)

\SetWidth{2.}
\LongArrow(0,20)(140,20)
\SetWidth{1.}
\Text(80,6)[l]{beam remnant}
\Text(143,20)[l]{u}

\LongArrow(0,20)(50,50)
\Text(52,50)[bl]{$\overline{\mathrm{c}}$}

\LongArrow(0,20)(60,36)
\Text(62,38)[bl]{$\overline{\mathrm{D}}$}
\SetWidth{4}
\qbezier(50,50)(50,30)(140,20)

\end{picture}

\begin{picture}(170,55)(0,0)

\SetWidth{2.}
\LongArrow(0,20)(160,20)
\SetWidth{1.}
\Text(80,6)[l]{beam remnant}
\Text(164,20)[l]{ud}

\LongArrow(0,20)(50,50)
\Text(52,50)[bl]{c}

\LongArrow(0,20)(70,32)
\Text(72,34)[bl]{D}
\SetWidth{4}
\qbezier(50,50)(60,23)(160,20)

\end{picture}
\end{center}
\caption{The ud diquark of the proton remnant has, at least in the average,
a larger energy/momentum than the lone u quark. Because the diquark is in a
colour singlet with the c-quark, in average, we will have
$\langle y_\mathrm{D} \rangle > \langle y_c \rangle$ and $\langle y_\mathrm{D}
\rangle > \langle y_\mathrm{\overline{D}} \rangle$ for $y_\mathrm{beam remnant} > 0$.}
\label{fig:beam}
\end{figure}
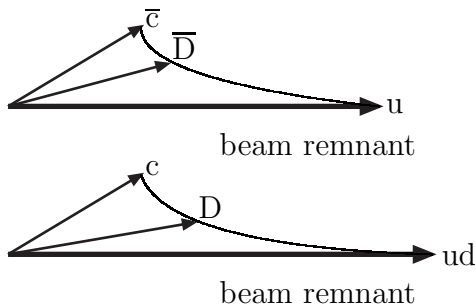

Beam remnant drag is a different, less obvious, source of asymmetry.
In an independent fragmentation scheme the rapidity of a charm quark
is preserved, on the average, during fragmentation.
This is not necessarily so in string fragmentation, where the produced hadron
inherits energy and momentum from both endpoints of the string. The resulting hadron will then
have a rapidity that, on the average, is shifted in the direction of the other end of the string.
An example with a proton beam remnant is shown in fig.~\ref{fig:beam}.

This is why the asymmetry for negative $\xF$ in fig.~\ref{fig:exp} is negative, even
though $\D^-$ is a leading particle also in this region. This behaviour is strongly
dependent on the choice of beam remnant distribution functions, see~\cite{ENTS}.

\section{Applications}

We have tuned the parameters of the model from a consideration of fixed target charm
production data. Other applications include any collision between hadron-like particles e.g.
$\gamma\p$, pp or $\p\pbar$. This sections gives just a few examples.

\begin{figure}
\begin{center}
\mbox{\epsfig{file=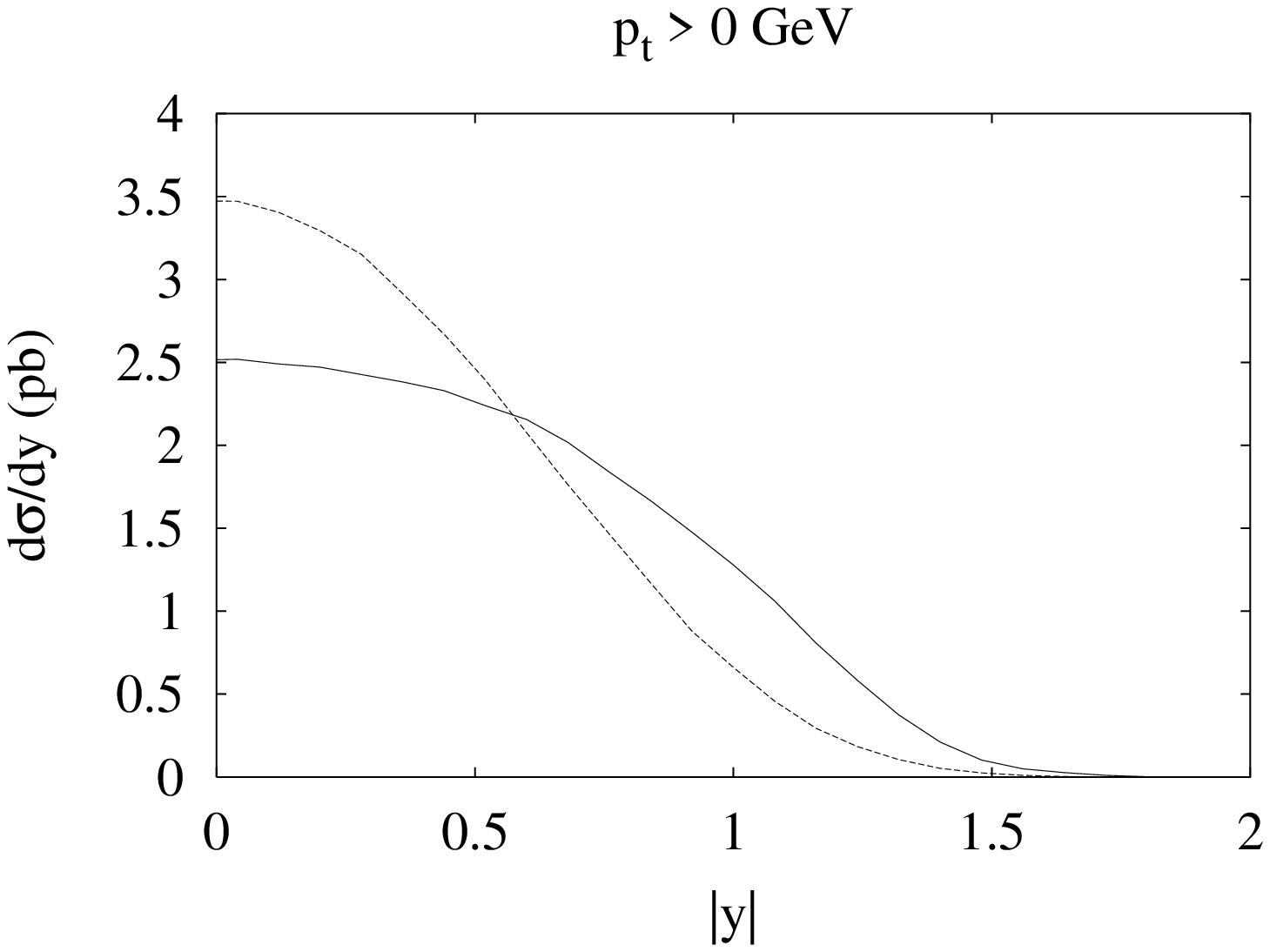, width=79mm}}
\mbox{\epsfig{file=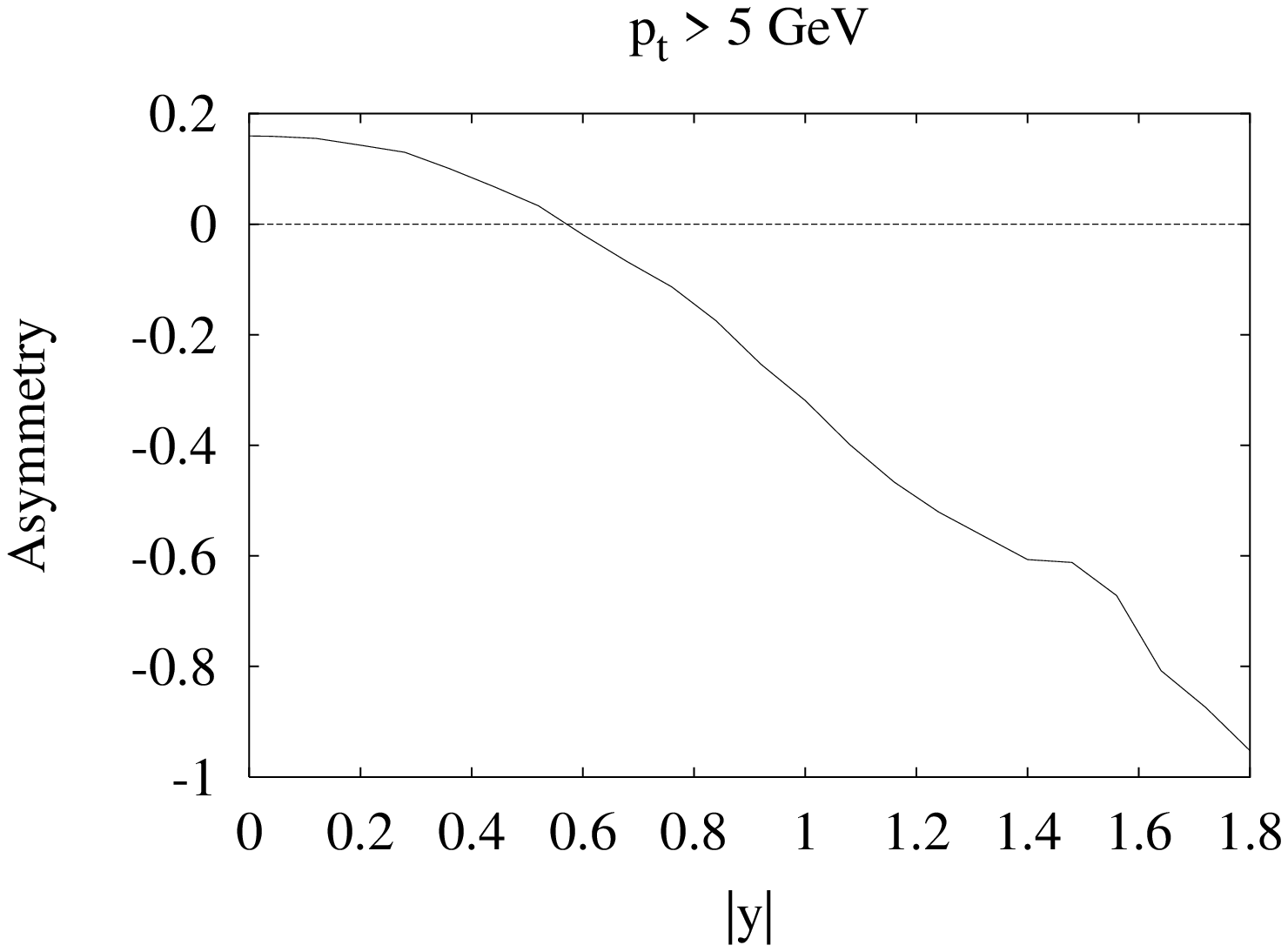, width=79mm}}
\mbox{\epsfig{file=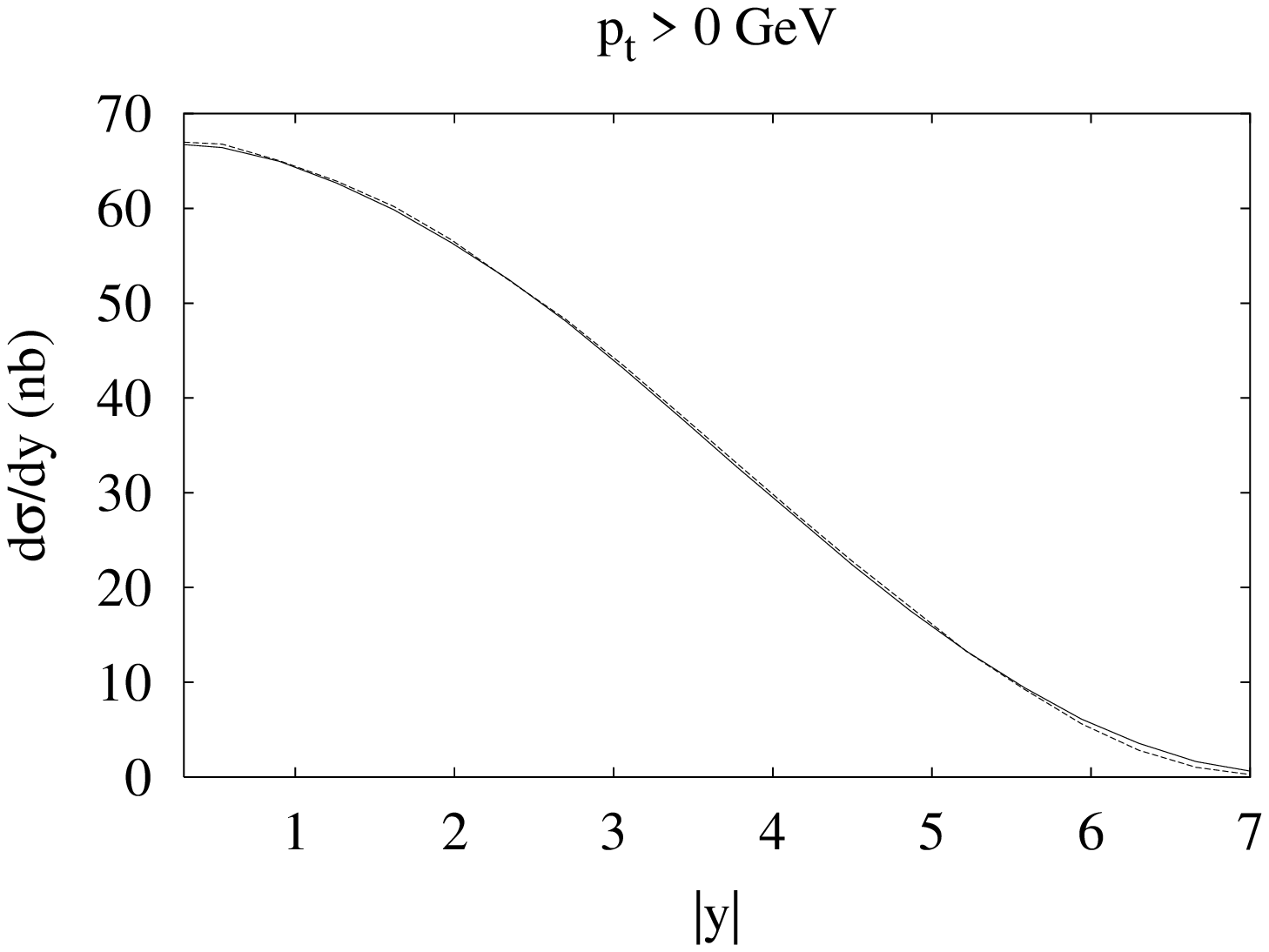, width=79mm}}
\mbox{\epsfig{file=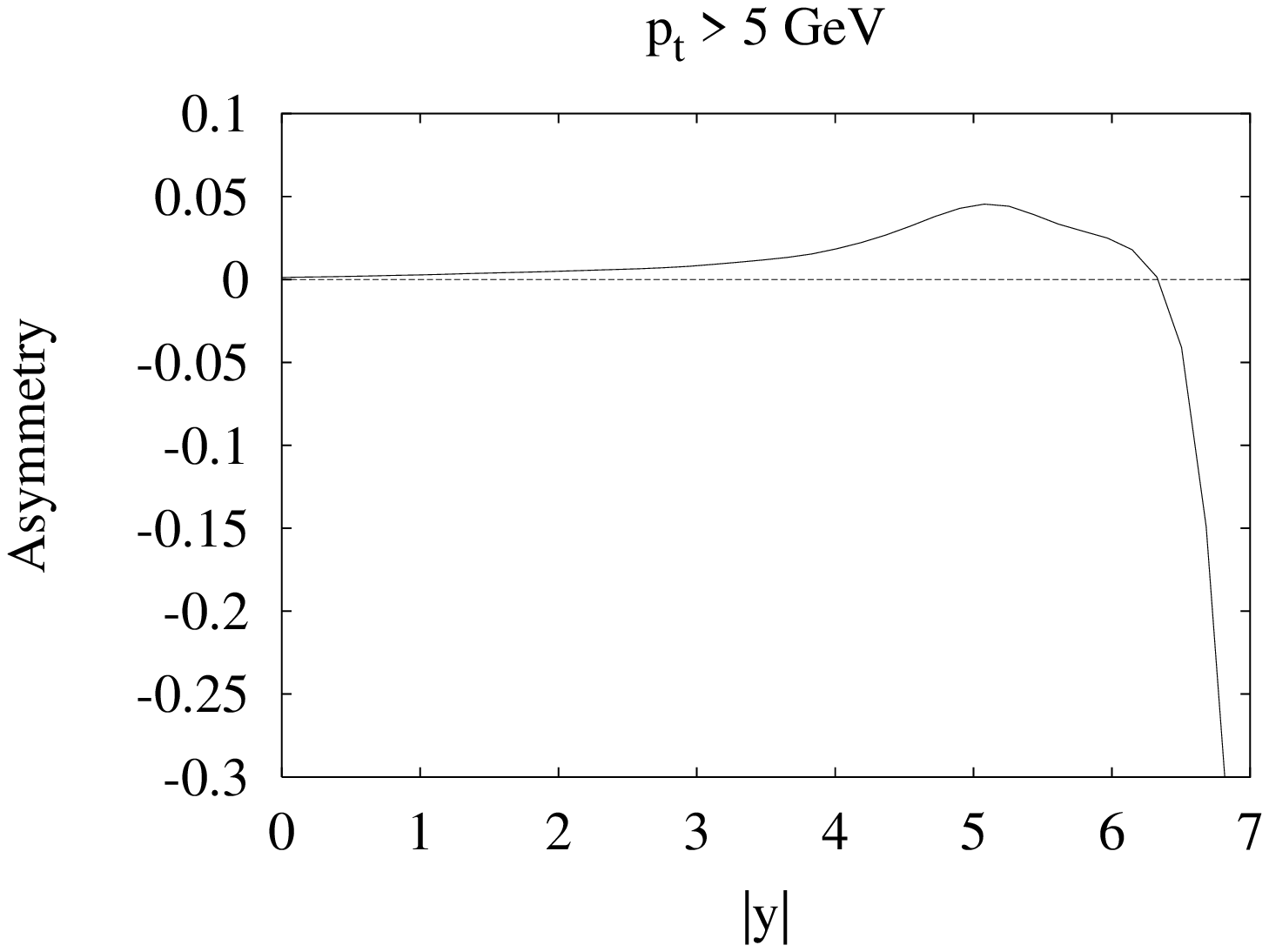, width=79mm}}
\end{center}
\caption{$\B^0$ (dashed) and $\Bbar^0$ (full) distribution and the resulting asymmetry,
$A = (\sigma(\B^0) - \sigma(\Bbar^0))/(\sigma(\B^0) + \sigma(\Bbar^0))$
at HERA-B (top) and the LHC (bottom).}
\label{fig:collasym}
\end{figure}

Asymmetries in the $\B^0\overline{\mathrm{B^0}}$ system at pp colliders must be considered as a background
to asymmetries from CP-violation~\cite{CPviol}. Fig.~\ref{fig:collasym} shows the bottom asymmetry
as a function of rapidity for HERA-B and LHC energies (pp  at 40 GeV and 14 TeV respectively).
At HERA-B the effect is as large as in fixed-target charm experiments and a measurement
of the effect would be feasible. At 14 TeV on the other hand, effects are at the
1\% level and will be hard to see. Because most strings at the LHC are very massive,
collapses will be rare. Only the drag effect for large rapidities is of any importance,
as can be seen from the figure. At higher energies, higher order effects will become
important -- e.g. at the LHC, less than half of the total heavy quark  cross section comes
from the leading order fusion processes included in fig.~\ref{fig:collasym}.

\begin{figure}
\begin{center}
\mbox{\epsfig{file=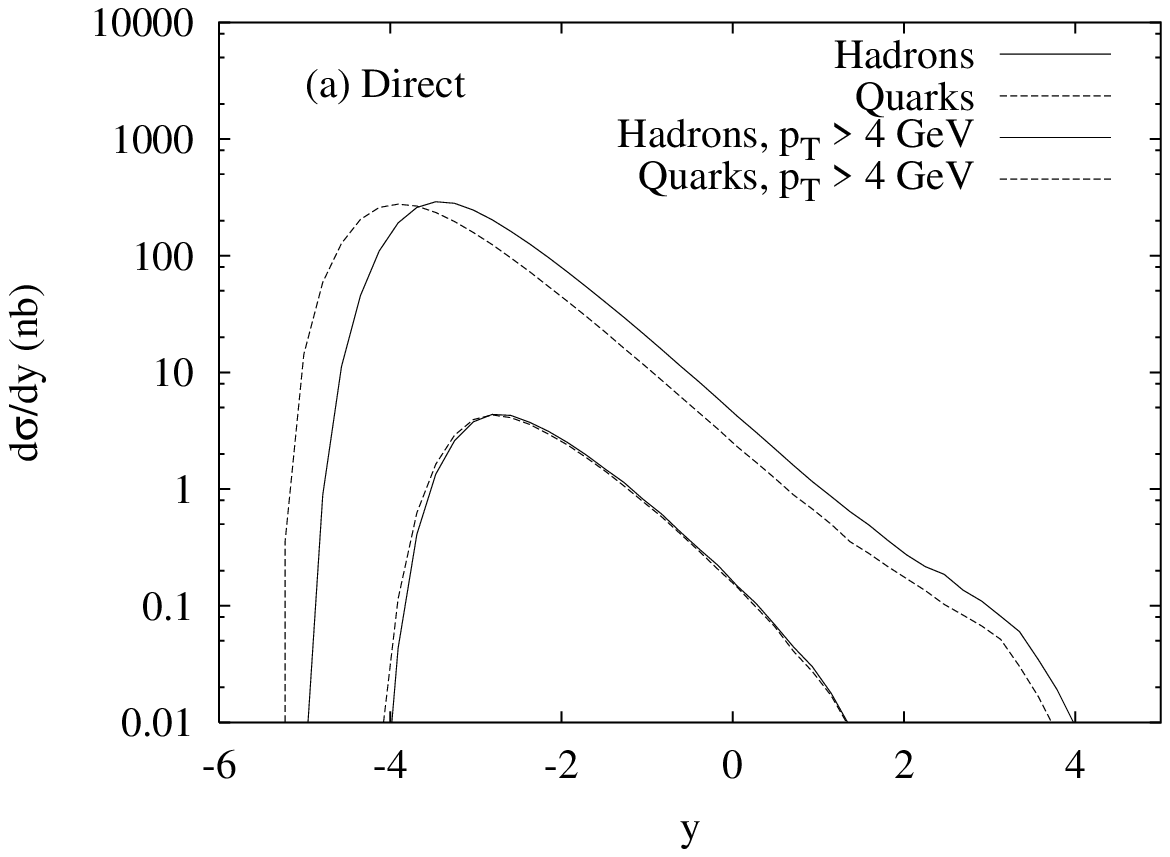, width=79mm}}
\mbox{\epsfig{file=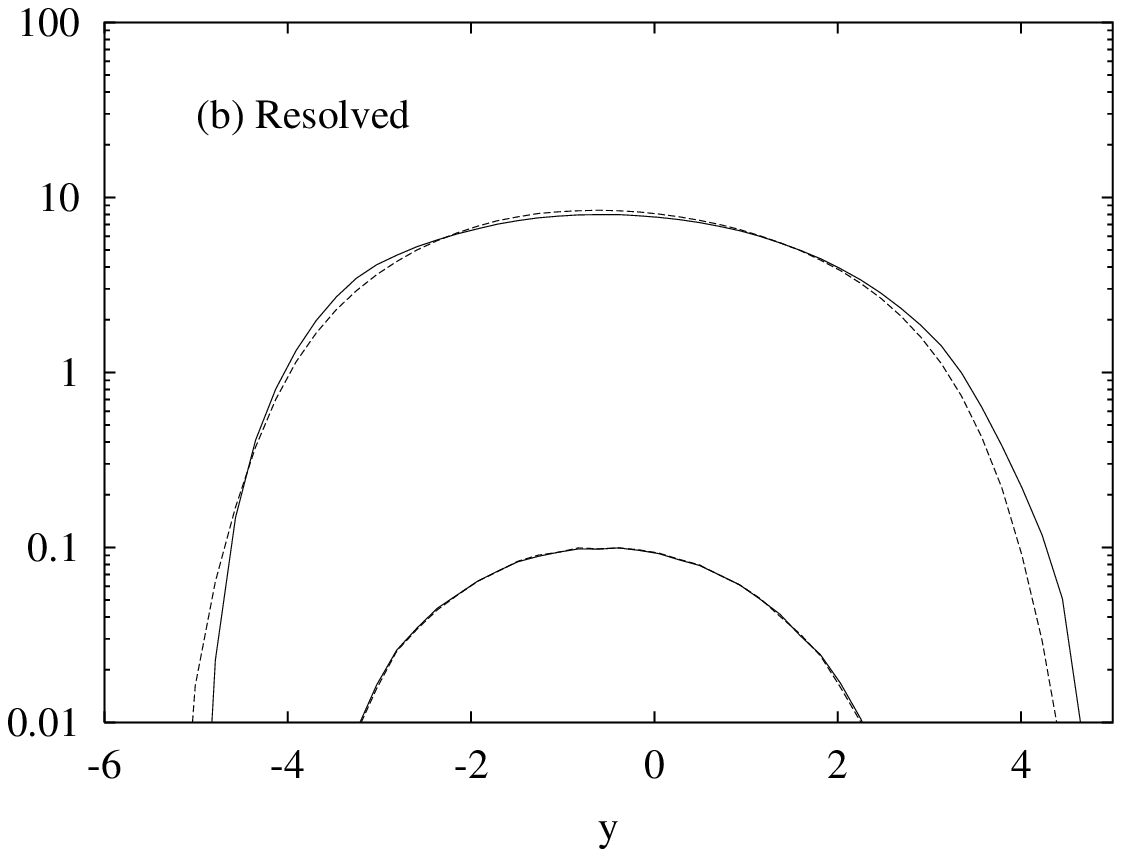, width=79mm}}
\mbox{\epsfig{file=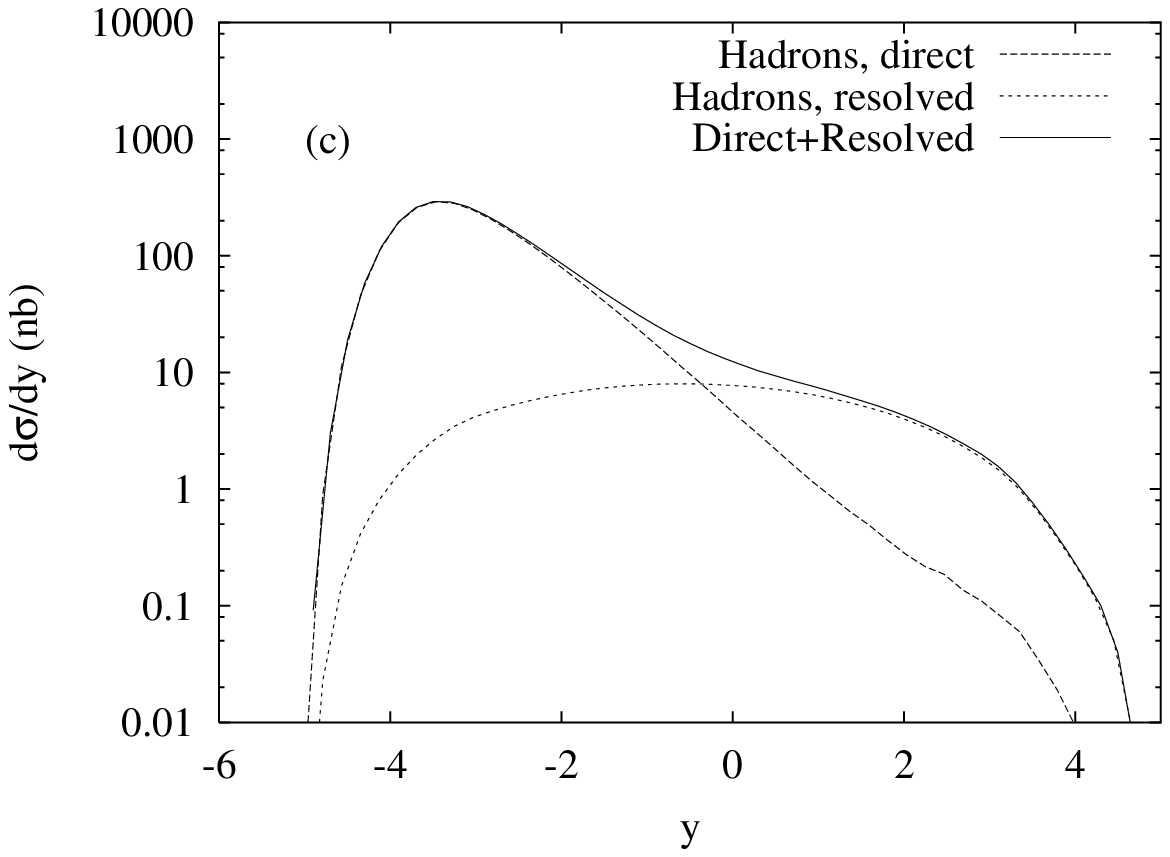, width=79mm}}
\mbox{\epsfig{file=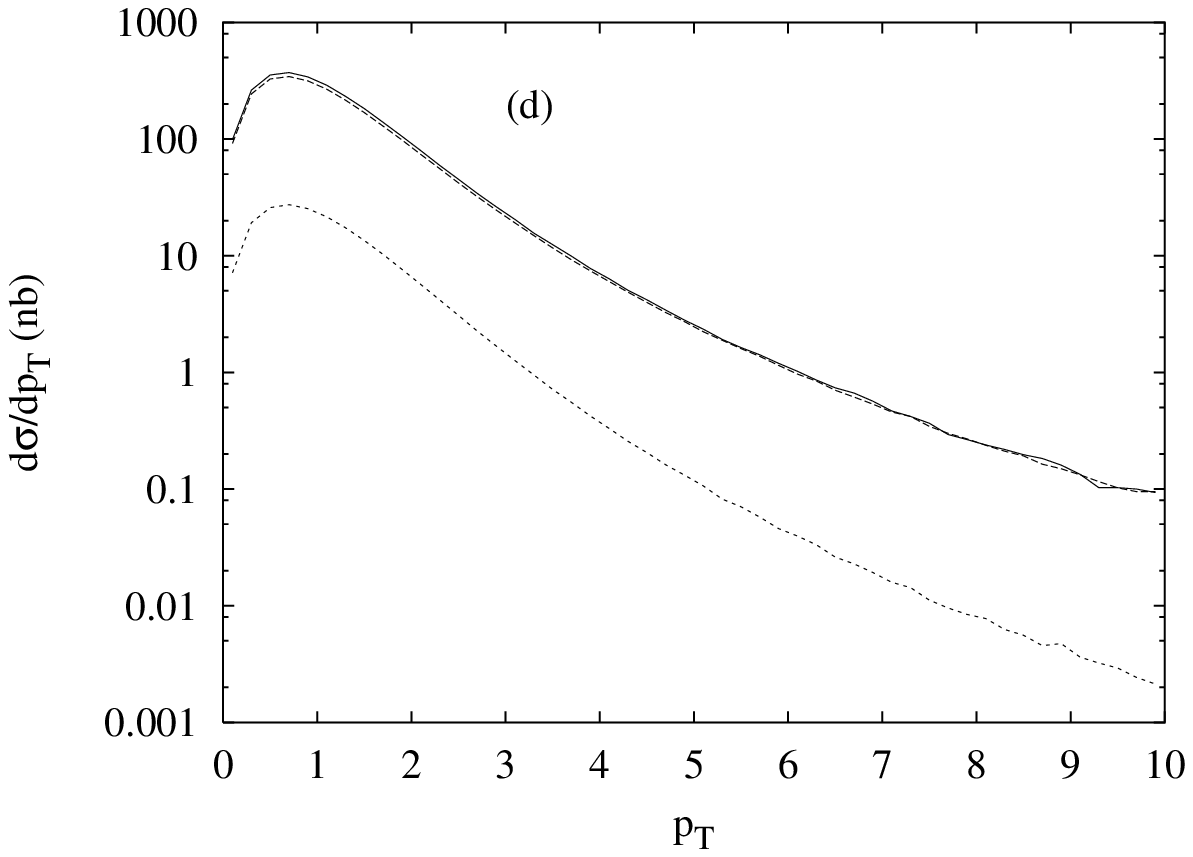, width=79mm}}
\end{center}
\caption{Charm cross section for direct and resolved photons with different $\pt$ cuts.}
\label{fig:photo}
\end{figure}

We also show some examples of charm photoproduction in $\gamma\p$ collisions. In this case
the study is further complicated because of the photon structure, which can be classified as
direct or resolved. In the direct case, the photon interacts as a point particle in
the hard scattering subprocess. If the photon has fluctuated into a $\q\qbar$ pair
it can interact as a meson and therefore you need to know the parton distribution of the photon.
These two classes of events give rise to fairly different event structures,
see fig.~\ref{fig:photo}. The drag effect shifts the hadron spectrum in the direction
of the proton beam remnant.

\section{Summary}
We have studied some non-perturbative effects in heavy quark production within the
framework of the Lund string fragmentation model. We find some agreement with data
if specific parameters are used. We also apply the model to pp and $\gamma\p$ physics
and arrive at predictions of bottom-asymmetries at HERA-B and the LHC.

\end{document}